\documentclass{article}
\usepackage{graphicx} 
\usepackage[
backend=biber,
style=nature,
]{biblatex}
\usepackage[parfill]{parskip}
\usepackage{authblk} 
\usepackage{amsmath}
\usepackage[left=1in,
  right=1in,
  top=1in,
  bottom=1in]{geometry} 
  
\usepackage{booktabs}
\usepackage{color}
\usepackage{url}
\usepackage{hyperref}

\begin{document}

\title{Scaling Laws for Function Diversity and Specialization Across Socioeconomic and Biological Complex Systems}


\author[a,b]{Vicky Chuqiao Yang}
\author[c]{James Holehouse}
\author[c,d,e]{Hyejin Youn}
\author[c]{Jos\'{e} Ignacio Arroyo}
\author[c]{Sidney Redner}
\author[c, f]{Geoffrey B. West}
\author[c]{Christopher P. Kempes}

\affil[a]{MIT Sloan School of Management, Massachusetts Institute of Technology, Cambridge MA, USA}
\affil[b]{Institute for Data, Systems, and Society, Massachusetts Institute of Technology, Cambridge MA, USA}
\affil[c]{Santa Fe Institute, Santa Fe, NM, USA}
\affil[d]{Graduate School of Business, Seoul National University, Seoul, South Korea}
\affil[e]{Northwestern Institute on Complex Systems, Evanston, IL, USA}
\affil[f]{London School of Economics and Political Science, London, United Kingdom}
\date{}

\maketitle
\section*{Abstract}
Function diversity, the range of tasks individuals perform, and specialization, the distribution of function abundances, are fundamental to complex adaptive systems. In the absence of overarching principles, these properties have appeared domain-specific. Here, we introduce an empirical framework and a mathematical model for the diversification and specialization of functions across disparate systems, including bacteria, federal agencies, universities, corporations, and cities. We find that the number of functions grows sublinearly with system size, with exponents from 0.35 to 0.57, consistent with Heaps' Law. In contrast, cities exhibit logarithmic scaling. To explain these empirical findings, we generalize the Yule-Simon model by introducing two key parameters: a diversification parameter that characterizes how existing functions inhibit the creation of new ones, and a specialization parameter that describes how a function's attractiveness depends on its abundance. Our model enables cross-system comparisons, from microorganisms to metropolitan areas. The analysis suggests that what drives the creation of new functions depends on the system’s goals and structure: federal agencies tend to ensure comprehensive coverage of necessary functions; cities tend to slow the creation of new occupations as existing ones expand; and cells occupy an intermediate position. Once functions are introduced, their growth follows a remarkably universal pattern across all systems.

\section*{Significance Statement}
Diversification and specialization are central to complex adaptive systems, yet overarching principles across domains remain elusive. We introduce a general theory that unifies diversity and specialization across disparate systems, including microbes, federal agencies, companies, universities, and cities,  characterized by two key parameters. We show from extensive data that function diversity scales with system size as a sublinear power law—resembling Heaps’ law—in all but cities, where it is logarithmic. Our theory explains both behaviors and suggests that function creation depends on system goals and structure: federal agencies tend to ensure functional coverage; cities slow new function growth as old ones expand, and cells occupy an intermediate position. Once functions are introduced, their growth follows a remarkably universal pattern across all systems.

\section*{Introduction}
Fundamental to complex adaptive systems---encompassing biological organisms, human organizations, and urban areas---is the range of functions (diversity) and their relative abundance (specialization). Diversity,  the number of unique sub-components within a system, indicates a broad range of abilities and capabilities for organisms or organizations to tackle complex challenges, hedge against risks, and thus adapt to complex environments (see Table \ref{tab:exponents}  for some examples)\cite{page2008diversity, whittaker1972diversity, quigley1998diversity, youn2016diversity, bettencourt2014diversity, Kline2010}. 
As these systems mature, develop, and expand, their function diversity grows, reflecting an increase in adaptability and resilience \cite{simpsonindex1949, page2008diversity, whittaker1972diversity, Kogut1992, quigley1998diversity}.
Remarkably consistent empirical regularities are observed in both biological and socio-economic systems, despite their distinct environmental conditions and evolutionary trajectories. Examples include the species-area curve---the empirical regularity for how the number of plant and animal species increases with land area \cite{whittaker1972diversity}---and the scaling of microbial richness with community size \cite{locey2016scaling}. A similar pattern is observed in urban areas---the range of occupations and business types is strongly associated with population \cite{youn2016diversity, bettencourt2014diversity, Hong2020}. 

While diversity provides a wide foundation of unique capabilities, specialization refines these capabilities for targeted functions, concentrating on a select few functions or skills to increase performance and competitive edge \cite{Perloff1960, Kim1995}.  For example, certain flowers adapt by developing nectars tailored to their specific pollinators, universities may concentrate on specific research areas like quantum computing, and cities may become hubs for certain sectors such as financial services. Specialization is manifested in the abundance distribution of functions---whether components or resources are concentrated in a few functions or evenly spread across many. Empirically, specialization has been quantified using indices like the Gini coefficient, location quotient, and entropy measures, all of which capture information in the abundance distribution \cite{frank2018, Vernon1991, Palan2010, Kim1995}. Specialization within the diversification process provides insights into how systems efficiently allocate resources and stand out in competitive environments \cite{page2010diversity}.  

Indeed, different systems pursue distinct strategic choices, objectives, and impacts under different specific scopes and subject matters. Accordingly, studies of diversity and specialization have predominantly been siloed within isolated disciplinary boundaries, thus limiting the possibility of a comprehensive understanding. The existing literature in both economics \cite{beaudry2009s, imbs2003stages} and biology \cite{macarthur1967limiting,cracraft1985biological,mahler2013exceptional,day2016specialization} often presents diversification and specialization as opposing poles of a linear spectrum. While this perspective is valid, especially in the context of finite resource allocation, it does not universally hold. For instance, research has demonstrated that introducing more minor competitors into a market can, paradoxically, reinforce the dominance of the major players \cite{chu2018theory}. This indicates that diversification and specialization may be more accurately understood as a two-dimensional problem. In addition, research efforts have often focused on specifics, ranging from the effects of team diversity \cite{page2008diversity, Modi2025} to the ramifications of organizational diversity \cite{palich2000curvilinearity, horwitz2015working, miliken1996diversity}, labor specialization \cite{becker1992divisionlabor, Hosseinioun2025}, cellular heterogeneity \cite{elowitz2002stochastic,holehouse2020stochastic,altschuler2010cellular}, or economic diversity and complexity \cite{hausmann2014atlas, hidalgo2009building}. While these studies are invaluable within their respective fields, findings are often limited to a particular organizational type or the impact of single variables. This segmented approach runs the risk of missing out on identifying overarching principles and theories that underlie the processes of diversification and specialization across a variety of systems. 

Thus, there is a need for a comprehensive framework to study diversification and specialization across complex systems. Classically there have been two separate approaches to characterizing commonalities in complex systems: the distributional approach \cite{barabasi1999emergence,broido2019scale} and the feature scaling approach \cite{heaps1978information,west2018scale, west1999origin, bettencourt2007growth}. The first tries to connect generative mechanisms to the shape of abundance and probability distributions, the second connects mechanisms to how aggregated properties scale with the overall size of a system often in the form of a power law $Y = Y_0 N^{\beta}$, where $Y$ is an aggregated quantity, $N$ the system size, and $Y_0$ a normalization constant. Power-law scaling of aggregated properties is widely argued to be the product of adaptive systems evolving under fixed constraints that yield particular optima \cite{west2018scale, west1999origin, bettencourt2007growth}. Distributional regularities are argued to stem from many processes including self-organized criticality \cite{bak1987self, bak1988self}, simple growth and attachment rules leading to power-law degree distributions in networks \cite{barabasi1999emergence}, or sampling processes in, for example, ecological systems \cite{hubbell2011unified,harte2011maximum}.

Urban systems provide a rich example of how feature scaling relationships have proven helpful for building scientific understanding of complex, highly variable systems. Systematic scaling laws linking socioeconomic outputs and infrastructure costs to population size have inspired models proposing mechanisms including social network effects \cite{bettencourt2013origins, yang2019modeling}, geometric and spatial constraints \cite{molinero2021geometry}, and mobility patterns \cite{ribeiro2017model}. Although the scaling exponents of different urban quantities can vary, such differences can arise from common generative principles \cite{yang2019modeling}. Studies have also examined diversity in urban systems using this framework. While new economic functions often originate in large urban centers and subsequently diffuse to smaller ones \cite{pumain2006evolutionary, hong2020universal}, the diversity of business establishments and occupations exhibits striking regularities across sizes \cite{youn2016diversity, bettencourt2014professional}. 

Beyond urban areas, scaling laws have also been used for cross-system studies. A well-studied regularity is Heaps’ law, an empirical pattern originally observed in linguistics in how the number of unique words in a corpus scales with its length, showing a sublinear power law \cite{heaps1978information}. Similar scaling relationships were later observed in analyses from books in various languages to musical compositions to genomes—using appropriate diversity metrics, they have consistently identified power-law  scaling for how number of unique elements scale with size, with exponents between 0.35 and 0.60 \cite{mazzolini2018heaps, tria2014dynamics,benz2008discovery,tettelin2008comparative,park2019large,serra2021heaps}. The robustness of such empirical scaling laws in novelties within human artifacts has motivated theoretical efforts to uncover their generative mechanisms. Heaps’ law is closely related to abundance distributions, as it can be formally derived from Zipf’s law \cite{van2005formal}. Two main classes of explanations have been proposed---one emphasizing how existing elements open adjacent possibles that enable novel elements to emerge \cite{tria2014dynamics, loreto2016dynamics}, and another focusing on how early commitments constrain subsequent possibilities, limiting what can be added later \cite{corominas2015understanding}. Together, these findings suggest that scaling regularities may reflect universal principles governing how complex systems diversify and specialize as they grow. This conceptual foundation motivates our empirical and modeling analysis of how functional diversity and specialization scale with system size across biological and socioeconomic domains.


In this study, we investigate function diversity and specialization in a broad range of socio-economic and biological systems, including bacteria, companies, government agencies, universities, and cities. We explore the interconnected nature of diversification and specialization across a wide range of systems by asking: Do various biological and social systems display common patterns of function diversity and abundance, and to what extent? What general governing principles could determine the function diversity and specialization of complex adaptive systems?  We first measure the range (diversity) and concentration of functions (abundance distributions) within each system. We identify functional forms for each attribute to compare disparate systems. We then propose a general model for the growth of function diversity and abundance, explaining both the regularities and differences observed in the empirical data. 

Our cross-system perspective recognizes that biological and social systems exhibit common patterns of complexity, adaptation, and evolution. Certain principles, like network dynamics \cite{krapivsky2001organization} and feedback loops \cite{alon2007network}, are common across complex systems.  Exploring common governing principles across systems may reveal underlying universal patterns that govern complex systems in general. This view is supported by several prior studies that have found commonalities in related domains, such as in the distribution of business types in cities \cite{youn2016diversity, bettencourt2014diversity}, and the abundance fluctuations across biological, ecological, and urban systems \cite{george2023universal}.

\section*{Results}
\subsection*{Empirical Results}

\begin{figure}[tb]
    \centering
    \includegraphics[width = 1\textwidth]{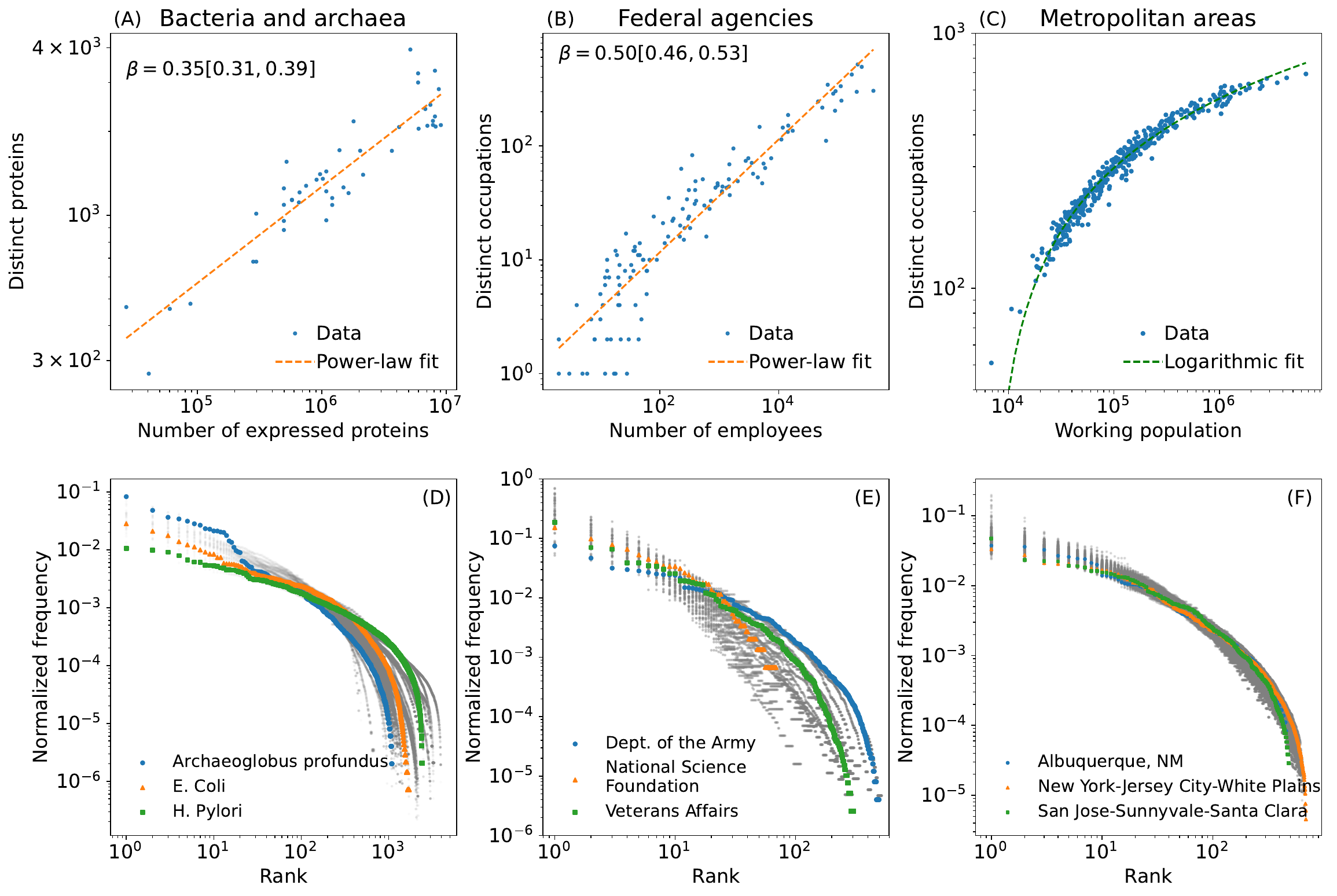}
    \caption{The number of distinct functions versus system size (top row) and rank-frequency distribution of function abundance (bottom row) in several complex adaptive systems. (A,D) Bacterial and archaea cells, (B,E) US federal agencies, and (C,F) US metropolitan statistical areas (MSA). The dependence of the number of functions on size measure $N$ in (A) and (B) can be approximated by the power law, $D \sim N^\beta$.  Also shown is the best-fit scaling exponent, $\beta$, and its $95\%$ confidence interval in the square brackets. (C) In contrast, logarithmic scaling occurs for MSA's, $D\sim \log N$. In (D--F), grey dots represent all available data points, while select entities are highlighted in color.}
    \label{fig:divScaling}
\end{figure}

\begin{table}[tb]
    \caption{Summary of scaling statistics for function diversity in various biological and social complex adaptive systems}
    \centering
    \begin{tabular}{p{2.5cm}|p{2.8cm} p{2.8cm} l l l l}
    \toprule
         &  Size measure&Function diversity measure&$\beta$ &95\% CI & R-sq  & Obs   \\\midrule
         Bacteria \& archaea & Total expressed protein & Distinct expressed proteins& 0.35&[0.31, 0.39] & 0.86 & 47 \\ \hline
          Norwegian companies& Number of employees& Distinct occupations& 0.46&[0.45, 0.46] & 0.88 & 3,191\\ \hline
 US federal agencies& Number of employees& Distinct occupations& 0.50&[0.46, 0.53] & 0.87 & 125 \\\hline
  US universities, associate level& Number of faculty& Distinct academic programs& 0.52&[0.50, 0.55]  & 0.61 &1,069 \\ \hline
 US universities, bachelor level \& above& Number of faculty& Distinct academic programs& 0.57&[0.55, 0.59]& 0.75 & 1,328\\ \hline
 Cities& Working population& Distinct occupations&  \multicolumn{2}{c}{Logarithmic scaling}&0.94 &422\\ 
 \bottomrule
 \end{tabular}
    \label{tab:exponents}
\end{table}

We define a function as the kind of task an individual component within a system is capable of performing, and function diversity as the number of distinct task types present in the system. This definition is intentionally agglomerated to enable comparisons across structurally and mechanistically diverse systems, ranging from cells to cities. While the internal architectures and generative mechanisms of these systems differ—such as gene regulation in cells, hiring decisions in firms, and curriculum development in universities—we focus on their realized functional repertoires: the observed set of distinct roles or activities that contribute to system-level capabilities. 

In biology, proteins are often referred to as the workhorses of the cell because they carry out nearly all of its essential functions. Each type of protein performs a specific task. For example, enzymes catalyze biochemical reactions to produce particular metabolites; motor proteins move substances within the cell; transcription factors are regulatory proteins that control when other proteins are expressed, preventing the simultaneous production of proteins that could interfere with one another. These protein roles have analogs in many socioeconomic systems, such as factory production workers, delivery workers, and managers. Occupations in cities and firms capture the range of economic activities their components can carry out, while academic programs in universities reflect the range of instructional or research activities the institution can perform. Across domains, this functional repertoire reflects the system’s capacity to adapt, specialize, and perform complex tasks. By analyzing how function diversity scales with system size, we seek to uncover shared statistical regularities that may point to general principles. 

Our study analyzes empirical data from microorganisms to firms to metropolitan
areas, including 47 bacteria and archaea cell observations, 125 US federal government agencies, 3,191 observations of Norwegian companies, 2,397 US universities, and 422 US metropolitan areas.  For each system, we determine the diversity, the number of distinct functions $D(N)$ and the distribution of these functions' abundance as a function of system size $N$.

We define system size as the total number of individuals within a system. In social systems, this is simply the number of people in the system, with functions represented by occupations, which are categorized using standardized classifications (see SI for details on data and methods). In bacteria and archaea, key functional components responsible for metabolism and cellular processes are proteins, analogous to workers in an organization. Accordingly, we measure system size in these biological systems as the total number of expressed proteins, while function is quantified as the number of distinct expressed proteins. Figure~\ref{fig:divScaling} illustrates how the range of distinct functions, $D(N)$, scales with the system size $N$. This scaling framework allows for direct comparisons across diverse systems, as demonstrated in previous studies utilizing scaling analysis \cite{west2018scale, bettencourt2007growth,west1999origin}. The data for bacteria and archaea are visualized in panel A, federal agencies in panel B, and metropolitan areas in panel C. Analysis of additional datasets, including Norwegian companies and universities, are summarized in Table \ref{tab:exponents}, with visualization of scaling relationships included in the SI. 

For most systems, this scaling behavior is well-captured by a power-law relation $D = D_0 N^\beta$, where $\beta$ is the scaling exponent, and $D_0$ is a constant. For bacteria and archaea, companies, federal agencies, and universities (both associate and bachelor level and above), the scaling exponents are sublinear, ranging between $0.35$ to $0.57$. This consistency is particularly striking in bacteria and archaea, given that each of the proteomes comes from organisms that have followed unique evolutionary trajectories, have different metabolisms, and occupy unique ecological niches. The sublinear nature of this scaling ($\beta < 1$) further implies that as systems grow in size, the increase in the number of distinct functions is less than proportional. This suggests a tendency for larger systems to reuse or replicate existing functions rather than continuously introducing new ones. 

Notably, this behavior closely resembles Heaps' Law, the sublinear power-law scaling for how the number of unique elements scale with size in entities ranging from books in various languages to musical compositions to genomes. While Heaps’ Law in human systems is usually associated with the accumulation of novelty in produced artifacts---such as vocabulary in texts or motifs in music---our findings suggest that this same empirical regularity extends to the structure of many human organizations themselves. In other words, the sublinear growth of diversity is not limited to what humans create, but also characterizes many agglomerations they form, such as companies and universities. This shift—from artifacts to collectives---may suggest a deeper principle underlying how complexity scales in systems shaped by human activity.

Among the systems analyzed, a notable exception is metropolitan areas. Unlike other datasets, the scaling of function diversity in metropolitan areas does not follow a power-law pattern but is better captured by a logarithmic relationship, $D =  b \log N + c$ , as shown in Figure \ref{fig:divScaling}(C) (in SI, we show this logarithmic scaling relationship is robust over time). This observation suggests that different mechanisms govern the growth of function diversity in metropolitan areas compared to other systems. Function diversity grows more rapidly with population in smaller cities but decelerates as cities become larger. Notably, the logarithmic scaling observed in metropolitan areas departs from Heaps’ Law. This deviation indicates that Heaps-like behavior is not ubiquitous, but instead contingent on structural factors that shape how systems diversify and specialize. If a fundamental underlying mechanism for diversity and specialization exists, it must be capable of explaining both scaling behaviors---under some conditions, Heaps-like behavior emerges, while in others, diversification and specialization follow alternative patterns.

This logarithmic scaling behavior carries two important implications for urban areas. First, if a city's population grows exponentially with time ($t$), i.e., $N(t) = N_0 e^{\alpha t}$, where $N_0$ is the population at time $t = 0$, and $\alpha$ is the growth rate, then function diversity is predicted to grow linearly in time, with $D = \alpha b t + D_0$, where $b$ is the parameter fitted from $D = b \log (N) + c$, and $D_0 = b \log (N_0) + c$ is a constant determined by the initial condition of the system. Then the growth rate of function diversity  ($dD/dt$) is given by the product, $\alpha\,b$, which is proportional to the population growth rate. For example, the growth rate $\alpha$ for US urban areas on average is $0.71\%$ in 2019 \cite{worldBankData} while the fitted parameter $b$ from data is $128$, which together predict that the growth rate for function diversity ($dD/dt$) is $\alpha\,b =  0.9$ occupations per year. Second, since most socio-economic metrics in urban areas, such as GDP, wages, and patent production, follow an approximate superlinear power-law relationship with population size \cite{bettencourt2007growth}, our findings suggest that function diversity serves as a key driver of recombinant growth \cite{Weitzman1998, youn2015recombinant}. As functional diversity expands, the combinatorial possibilities for innovation and productivity multiply, contributing to exponential increases in economic output. This underscores the central role of diversity in sustaining and accelerating economic performance in urban systems.

Lastly, we analyze the distribution of function abundance within each organization. Figure~\ref{fig:divScaling}(D), (E), and (F) illustrate the variation in the relative abundance of functions as a function of rank for bacteria and archaea, federal agencies, and metropolitan areas, respectively. Grey dots represent all data points, while select entities are highlighted in color for clarity.

Despite variation in the most prevalent functions across organizations—such as ``nurse'' and ``medical officer'' dominating in Veterans Affairs, while the National Science Foundation’s top occupations include ``miscellaneous administration and program'' and ``management and program analysis''---the rank-frequency distributions exhibit a consistent concave shape across all systems. This pattern suggests an exponential decline in function abundance with increasing rank. Notably, the distribution in metropolitan areas displays a striking level of uniformity compared to the more heterogeneous patterns observed in federal agencies and cellular systems. This consistency in urban areas aligns with previous research highlighting similar diversity patterns in business types across cities \cite{youn2016diversity, Hong2020}.

Our analysis reveals both commonalities and differences across systems. The commonality is manifested in the fact that within a given system (bacteria, companies, etc.), the range of function diversity generally grows as a sublinear power law with system size. This commonality suggests that while varying in goals, environment, and history, different organisms may share a similar underlying dynamic in how the number of unique functions grows with size. The differences across systems, however, are also important and manifested in two ways. First, the number of unique functions in cities scales differently from the other systems, suggesting the mechanisms that lead to diversity in cities may be fundamentally different from the other entities. Second, among the systems that exhibit power-law scaling, there are variations in the scaling exponents. Our analyses provide the first quantitative assessment enabling comparison across disparate systems.

These findings present both challenges and opportunities for a comprehensive theory that can account for both the observed commonalities and differences.
In particular, while existing models have typically predicted either power-law or logarithmic scaling of function diversity (e.g., \cite{loreto2016dynamics, youn2016diversity}), none have successfully integrated both patterns. In the next section, we propose a model that is capable of explaining the full spectrum of empirical observations related to function diversity and specialization. By identifying and incorporating key model parameters, our approach aims to shed light on the diverse mechanisms at play, offering a unified understanding of function diversity and specialization across systems.


\subsection*{Mathematical Model}

\begin{figure}[htb]
    \centering
    \includegraphics[width = 0.6\textwidth]{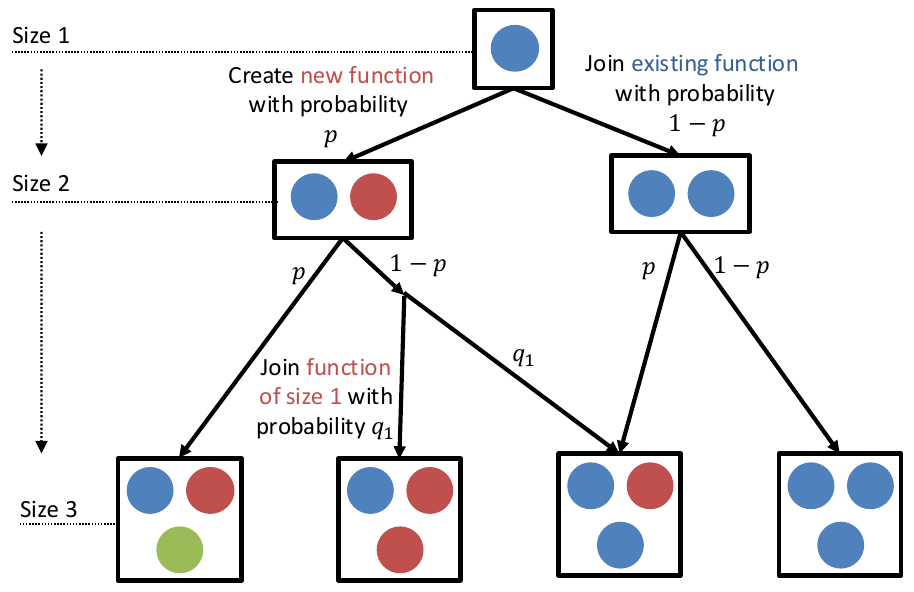}
    \caption{Conceptual diagram of the mathematical model. As each new individual joins the system, it either creates a new function with probability $p$, where $p$ depends on the existing abundance distribution; or joins an existing function with probability $1-p$. When it joins an existing function, the probability of joining a particular existing function with $k$ members, $q_k$, follows a generalized preferential attachment process.}
   \label{fig:model_cartoon}
\end{figure}

Our model extends the foundational principles of the Yule-Simon model \cite{yule1925ii,simon1955class}, to describe the growth dynamics of a system as new individuals join.  Within this framework, a new member that joins the system can lead to one of two scenarios: (i) the new member creates a new function (diversification, with probability denoted as $p$) or (ii) the newcomer assimilates into a pre-existing function (specialization, with probability $1-p$). For example, the individuals entering the system are: newly synthesized proteins in a bacteria; newly hired staff in a federal agency; and a new professional entering the labor market of a city.  
%
We assume Markovian dynamics throughout the model, which implies that the probability of adding a new function, or expanding an existing function, is only dependent on the present state of the abundance distributions. Figure \ref{fig:model_cartoon} visually summarizes this dynamic process.

Unlike the classic Yule-Simon model, which assumes a constant rate $p$ of new function creation, our approach introduces a dynamic framework. As a system adapts to a changing environment, new functions may become necessary. However, as the number of existing functions increases, it becomes more likely that the needed functionality is already present. Consequently, the rate of new function creation decreases. When the system contains all possible functions, this rate should approach zero. A simple functional form that would capture the decreasing tendency of $p$ is $p = p_0/D$. The constant $p_0$ represents effects from environmental factors that are assumed to be the same across systems of the same type. 

More generally, the rate of new function creation may not scale inversely with diversity alone, but instead depend on the system's internal structure, as captured by its rank-frequency distribution. A flexible parameterization that accounts for this is,
\begin{equation}\label{eq:p_2}
     p = \frac{p_0}{\sum_{i = 1}^D k_i^\theta} \;.
 \end{equation}
where $k_i$ denotes the number of individuals within function $i$. The parameter $\theta$ modulates the influence of function abundance on the creation of new functions. When $\theta = 0$, we recover the earlier linear dependency, $p = p_0/D$. For $\theta> 0$, functions with more individuals more strongly suppress the addition of new functions. For $\theta< 0$, the opposite holds. This generalized form allows for a wide range of dynamics to be modeled at the level of individual systems.

When a new individual joins an existing function, we consider the probability of the individual joining each function to be affected by the abundance of the existing functions. Having more individuals in a function can increase the likelihood of a new individual joining this function. In some scenarios, it may have the opposite effect because the existing abundance already adequately meets the system's demands. We can formulate this tradeoff as a nonlinear preferential attachment process---that the probability of an individual joining an existing function with $k$ individuals, $q_k$, is
\begin{equation} \label{eq:q}
    q_k =  \frac{ k^\gamma}{\sum_{i = 1}^{D} k_i^{\gamma}} \;,
\end{equation}
where $\gamma$ is a specialization parameter that indicates the extent to which large functions attract newcomers. This equation is similar to previous models of preferential attachment in physics \cite{barabasi1999emergence,krapivsky2001degree,krapivsky2001organization} and circular causation \cite{Myrdal1957}, positive feedback \cite{Arthur1990} or agglomeration \cite{Hoover1948} in economics. Our model introduced $\gamma$ as a parameter, ranging from 0 to 1, to control the strength of this positive feedback. For example, on the one hand, sublinear preferential attachment \cite{krapivsky2001organization}, that is, $\gamma < 1$, denotes diminishing returns in the effect of function abundance on joining an existing function. On the other hand, the cases $\gamma = 1$ indicate linear preferential attachment, and $\gamma>1$ corresponds to superlinear preferential attachment, which is extremely sensitive to the existing size of the function. 

In summary, our model contains two key parameters, $\theta$ and $\gamma$, that characterize the effect of \textit{diversification} and \textit{specialization}, respectively. While $\theta$ primarily determines the expansion of the function range, $\gamma$ primarily shapes the accumulation of individuals in functions.

\subsection*{Model predictions and categorization of systems by diversification and specialization parameters}

We simulate a system's growth based on Eqs.~\ref{eq:p_2} and \ref{eq:q} and compare the model's predictions with empirical data. The dynamics of the model are shown in Figure \ref{fig:model_cartoon}. Systems of various sizes are treated as points along the model’s growth trajectory, representing successive stages in the accumulation of individuals. This assumes that larger systems arise through the sequential addition of individuals to smaller ones. Under this framework, the dynamic model generates cross-sectional predictions that can be directly compared to observed patterns of diversity and abundance across systems of varying sizes. 

We then estimate the parameters $\gamma$ and $\theta$ for each system by comparing the function range and abundance generated by simulation with those from data on bacteria, federal agencies, and cities. This estimation is achieved by minimizing the Euclidean distance between normalized rank-frequency distributions in logarithmic-space using the adaptive differential evolution algorithm for optimization (see SI for methods). 

\begin{figure}[htb]
    \centering
    \includegraphics[width = 1\textwidth]{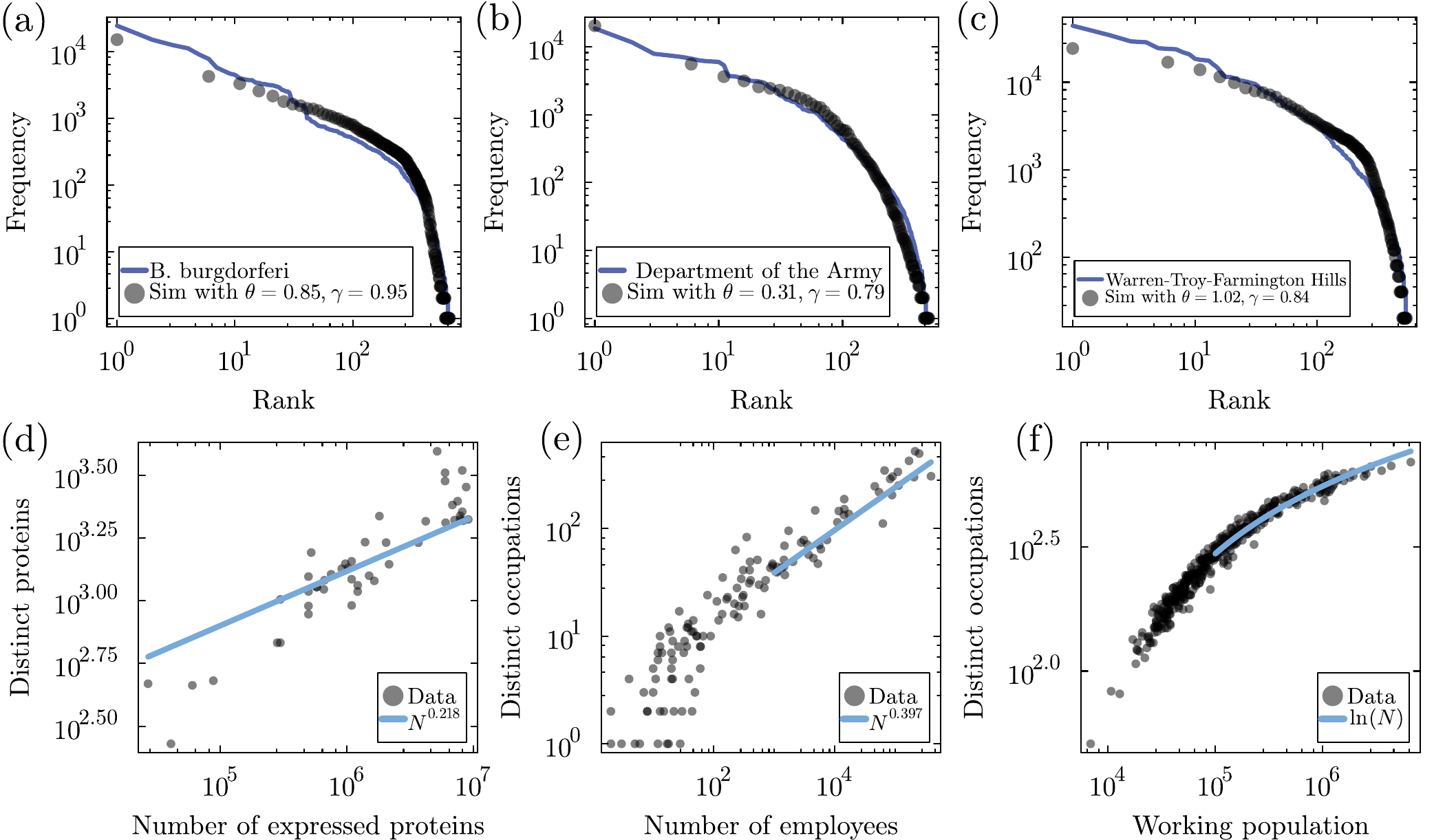}
    \caption{Summary of the model calibration results for the normalized rank-frequency distributions of (a) bacterial cells, (b) federal agencies and (c) cities, each for a given case of that system. The legends show the calibrated values of $\gamma$ and $\theta$, as well as the particular system in question. Each system was simulated using an initial condition that arose from one of the lesser size organizations in that data set (see SI for further details). Panels (d), (e), and (f) show the respective diversity plots from simulations given the mean values of $\theta$ across all calibrations for that system, starting from the initial condition size used for calibration. 
    }
    \label{fig:model_results_1}
\end{figure}

Figure~\ref{fig:model_results_1} (a-c) shows the model's predicted rank-frequency distributions are well aligned with three selected systems: Bartonella henselae (bacterial cell), the Army (federal agency), and Warren-Troy-Farmington Hills (urban area in Michigan).  See SI for more examples of rank-frequency distribution predictions. The bottom row of Figure \ref{fig:model_results_1} show their respective diversity relationship with those predicted by the model. Both results demonstrate good agreement across all three types of systems for both the rank-frequency distributions and the diversity scaling.

\begin{figure}[htb]
    \centering
    \includegraphics[width = 0.7\textwidth]{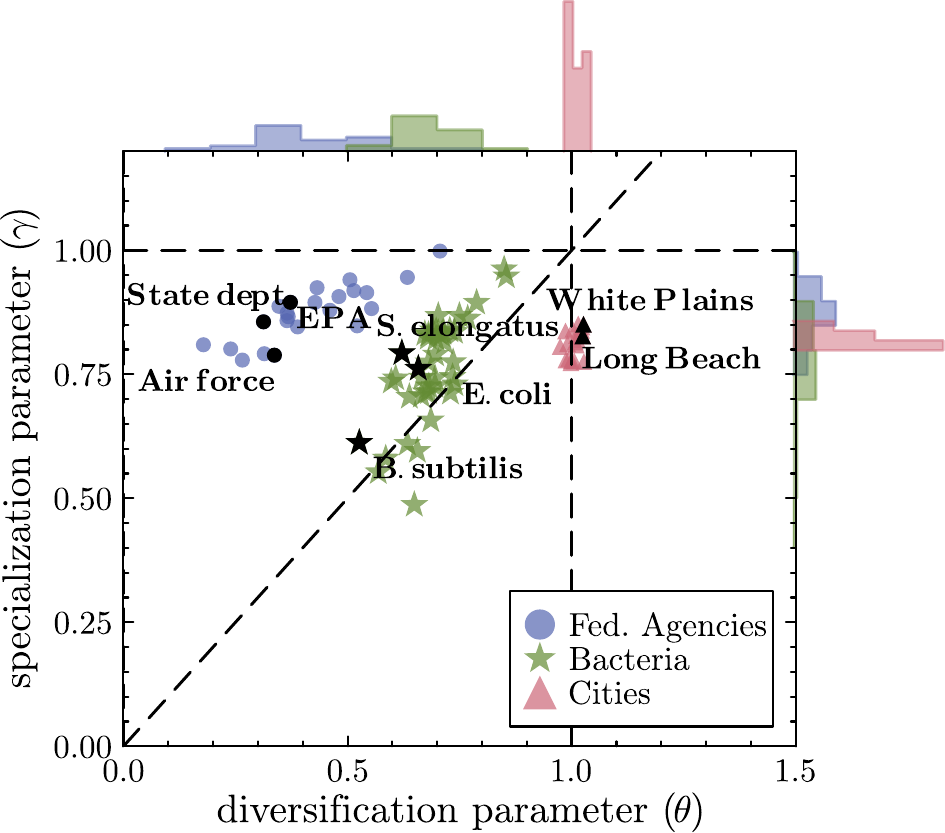}
    \caption{The parameter space of specialization ($\gamma$) and diversification ($\theta$) estimated for each system across different classes of systems. The vertical dashed line shows $\theta = 1$, the horizontal dashed line shows $\gamma = 1$, and the diagonal dashed line shows $\gamma = \theta$. We selected the federal agencies with $N>1000$ (40 instances), all of the bacterial data we had access to (46 species), and the largest twenty cities.}
    \label{fig:model_results_2}
\end{figure}

Figure~\ref{fig:model_results_2} summarizes our calibration results, mapping over a hundred cases onto the parameter space of $\theta$ and $\gamma$, where each marker represents the calibrated values of one entity, and marker shapes distinguish system type. Systems cluster by type, indicating distinct within-class behavior. The clustering patterns in Figure \ref{fig:model_results_2} reveal systematic differences in how functional specialization and diversification unfold across system types. Cities occupy a particularly narrow region of parameter space, centered around $\gamma \approx 0.8$ and $\theta \approx 1$, suggesting a highly consistent balance between diversification suppression and sublinear growth of existing functions. This tight grouping is likely associated with the uniformity in cities' abundance distributions compared to other entities (Fig.~\ref{fig:divScaling}(F)). In contrast, bacteria and federal agencies span a much broader region of the parameter space, likely reflecting differences in internal constraints and evolutionary or administrative heterogeneity. Together, these results indicate that cities are not only distinct but also unusually consistent in their system-level patterns compared with other entities.

At the same time, patterns emerge across systems along both $\theta$ and $\gamma$, revealing shared mechanisms that operate both within and across system classes. For instance, bacteria, federal agencies, and cities exhibit parallel trends in the specialization parameter $\gamma$, which captures how the abundance of existing functions shapes subsequent growth. All three systems consistently show positive $\gamma$ values, but predominantly sublinear (below 1), indicating a general tendency toward diminishing returns in specialization. Nevertheless, the variation in $\gamma$ is wider for federal agencies and bacteria than for cities, suggesting that urban systems may exhibit a more uniform—and possibly universal—specialization pattern. 

\begin{figure}[htb]
    \centering
    \includegraphics[width = 0.7\textwidth]{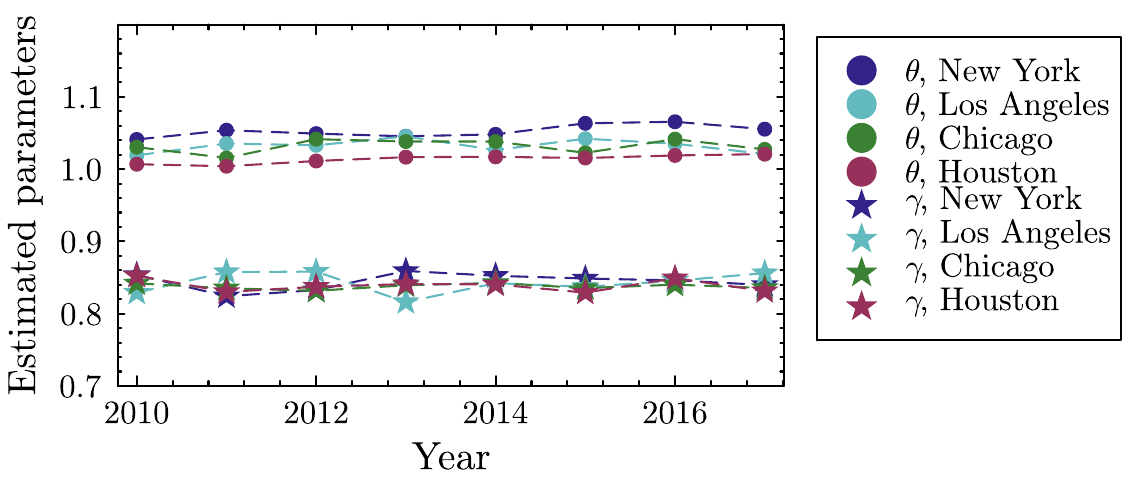}
    \caption{Calibrated parameters ($\theta$ and $\gamma$) for the four largest metropolitan areas in the longitudinal dataset, by year. Parameter values show no systematic temporal trends.}
    \label{fig:results_over_time}
\end{figure}

To test whether the parameter patterns observed for cities are robust to temporal and contextual variation, we repeated the calibration across multiple years for the four largest urban areas (Figure \ref{fig:results_over_time}). The results show no systematic temporal trends, despite population of these cities having changed significantly over this time period---indicating that both $\theta$ and $\gamma$ remain stable over time. We also perform robustness tests and demonstrate the stability of these parameters to data noise, initial conditions, and alternative assumptions (see SI).


While the specialization parameter, $\gamma$, exhibits common patterns across systems in terms of its mean value, the diversification parameter, $\theta$, differs significantly across systems. It is notably higher for cities than for federal agencies and cells. A higher $\theta$ value suggests the existence of a highly abundant function suppresses new function creation.  This is in line with the observation that large cities tend to become hubs for certain industries \cite{balland2020complex}, such as San Francisco Bay Area for information technology and the Greater Boston area for biotech. A value of $\theta$ around zero implies that an increased range of existing functions, regardless of the size of the functions, contributes towards suppression of new functions. 
The differences in $\theta$ parameters can help explain the differences in scaling behavior observed. When $\theta = 1$, given that $\sum_{i = 1}^D k_i = N$, we can simplify Eq.~\ref{eq:p_2} to derive $p = p_0/N$. Since $p = dD/dN$, we arrive at the dynamical equation $dD/dN =  a D_{\text{max}}/N $. The solution of this equation takes the form of $D \sim \log (N)$, which recovers the logarithmic scaling. Similarly, in the case of $\theta = 0$, we have $\sum_{i = 1}^D k_i^0 = D$. This leads to $dD/dN \sim 1/D$, which implies the power-law scaling $D \sim N^{1/2}$. For bacteria, Figure \ref{fig:divScaling}(A) shows that diversification happens faster than in cities, but slower than in federal agencies, and correspondingly the mean calibrated value of $\theta$ is between 0 and 1.

Each class of systems occupies distinct regions of parameter space. First, the values of $\theta$ found for cities imply that diversity growth is most impeded when large functions are abundant. Second, for federal agencies, smaller $\theta$ values imply that functions more equitably impede diversity growth. Finally, for bacterial cells, an intermediate $\theta$ value implies an in-between level of diversity growth still impeded by large functions. Cities express the largest value of $\theta$ (approximately 1) that simultaneously allows for the maximization of concentration in a few functions while maintaining consistent rank-frequency distributions without leading to gelation behavior that occurs for $\gamma>1$ \cite{krapivsky2001organization}. 
In this sense, gelation provides a natural bound to function growth since it does not allow for the robustness or diversity necessary for organizations to survive complex environments.

By enabling cross-domain comparisons, we identify important commonalities and differences across complex systems. The barrier to introducing new functions varies across system types. Yet, once a function is added, it tends to grow following a common sublinear preferential attachment dynamics. The variations may stem from differences in system goals and structures. Federal agencies and bacterial cells function as cohesive organisms with well-defined boundaries, where adaptation occurs at the level of the whole entity. In federal agencies, the introduction of new occupations is typically decided in a centralized manner. Cities, on the other hand, lack unified objectives and resemble ecosystems: decentralized, and adaptation occurs at the level of the individuals. The introduction of new occupational categories is decided in a decentralized fashion. For organism-like systems, it is beneficial to develop a full set of functional capabilities to effectively adapt to environmental challenges. Thus, even when some functions dominate, there remains a drive to cover all essential bases. In cities, however, large functions may inhibit the emergence of others due to competitive or self-reinforcing dynamics. Once functions are introduced into these systems, the growth of function abundance follows a shared pattern across all systems: sublinear preferential attachment. This suggests that while the drivers of function \textit{creation} differ based on system architecture and goals, the \textit{growth} of functions once introduced follows a remarkably universal process.

\section*{Discussion}
We identify common patterns in the diversity and abundance of functions across a broad spectrum of biological and social complex systems. In systems such as bacteria, federal agencies, companies, and universities, function diversity can be approximated by a power-law relationship with size, characterized by sublinear exponents ranging from 0.35 to 0.57, depending on the system. This is consistent with Heaps' Law. However, cities differ significantly from these systems and exhibit logarithmic scaling in function diversity and greater commonality in abundance distributions. This empirical anomaly may stem from cities’ fundamentally different structures and goals. 

To reconcile these empirical patterns, we propose a general model for the dynamics of function diversity and abundance. The model is characterized by two fundamental parameters: a diversification parameter, which governs the expansion of the function range, and a specialization parameter, which grows the function concentration. We estimate both parameters from the data for bacterial cells, federal agencies, and cities. Notably, we observe consistent sublinear preferential attachment across all three systems, with generally small variations in the specialization parameter. However, the diversification parameter varies significantly across systems, suggesting different drivers for function range expansion. We emphasize that this universality lies at the level of generative mechanism rather than identical empirical outcomes. For simplicity, the model that we propose is fully Markovian, although future extensions to non-Markovian models with memory \cite{ghorbani2018gene,reed2023role} would be a beneficial addition to the literature, especially in the context of memory of organizational routines. The tight clustering of cities in parameter space implies a high degree of structural regularity in how urban systems specialize and diversify. This consistency supports the view that urban economies may be shaped by shared underlying mechanisms, as proposed in the urban scaling literature \cite{bettencourt2021introduction}. Extending this analysis to cities across different geographical and governance contexts would provide a useful test of the model’s generality.

Our framework offers a complementary alternative to several influential theories in biology and ecology that have sought to explain the origins and patterns of diversity. Classical models such as the Island Biogeography Theory \cite{macarthur2001theory}, the Unified Neutral Theory of Biodiversity and Biogeography (UNTBB) \cite{hubbell2011unified, serra2013neutral}, and Maximum Entropy \cite{harte2011maximum} approaches have all been pivotal in modeling biodiversity through the lens of ecological equilibria, spatial constraints, or information-theoretic principles. In particular, the UNTBB has offered a controversial but widely used null model, helping researchers benchmark observed diversity patterns across ecosystems and even genomes. Unlike these models, which are typically grounded in ecological or evolutionary narratives specific to biological systems, our approach derives from a generalized process-based model that incorporates both diversification and specialization dynamics, and it applies equally well to socio-economic systems such as cities and federal agencies. This broader scope enables cross-domain comparisons and identifies shared underlying mechanisms, while also capturing key differences—such as the logarithmic versus power-law scaling of diversity—in a unified parameter space.

A key to this cross-domain comparison is the definition of diversity. Our use of the term in this paper specifically refers to the number of unique functions in the sub-components within a system. This definition differs from entropy-based measures stemming from information theory, which have also been used as a diversity index in the biological and social systems literature. Entropy-based measures combine two properties of distributions—richness and evenness—into a single index. While valuable in many contexts, entropy-based measures have the limitation of possessing mathematical properties that do not align with theoretical or intuitive understandings of diversity in applied domains \cite{jost2006entropy}. In contrast, our diversity metric solely measures richness, the system’s range of unique capabilities. We explicitly separate the dimension of evenness, which we refer to as specialization, into a distinct second axis.

Our analysis relies on classification systems that delineate what constitutes a distinct function in each domain, which have limitations. For example, a protein's role can sometimes vary depending on the cellular context across species, most retain consistent biochemical functions that permit meaningful functional classification. Social systems exhibit similar variations within function categories: occupations are defined by shared core tasks even as their implementation varies across organizations. The ``chief executive'' occupation category, for example, contains leaders of both start-ups and multinational corporations, whose daily activities differ, yet they are grouped together because they share core tasks of strategic oversight and coordination. The protein–occupation analogy thus reflects commonality in functional organization across diverse contexts rather than a literal one-to-one correspondence. Admittedly, classification systems, especially those in social systems, are shaped by administrative needs. However, they are updated regularly to reflect evolving labor and knowledge structures, and are widely used in real-world economic and policy decisions. We use the most granular level of these taxonomies to approximate a system’s repertoire of task specializations. Though no classification is perfect, the persistence of scaling patterns across domains that use different classification schemes suggests that the underlying dynamics of functional diversification and specialization reflect deeper regularities in system organization despite limitations in the classification systems.

Although proteins can exhibit context-dependent roles across species, most retain consistent biochemical functions that permit meaningful functional classification. Social systems show a comparable pattern: occupations are defined by shared core tasks even as their implementation varies across organizations. A start-up founder and a corporate CEO, for example, differ in daily activities yet share core functions of coordination and strategic control, and are thus categorized under the same SOC category. The protein–occupation analogy thus reflects a common structure of functional organization across diverse contexts rather than a literal one-to-one correspondence

Our work provides a parsimonious framework for unifying observed patterns across domains, but naturally abstracts away many details. This opens several directions for extension. While our model accounts for cross-sectional differences in scaling behavior across system types, future work could extend this framework to incorporate temporal dynamics, considering the evolution of the diversification and specialization parameters ($\theta$, $\gamma$) over time at different stages in system development. Our proposed framework represents an effective process, but does not explicitly capture the causal dynamics through which new functions or components emerge. Future work could extend this framework by explicitly modeling the structural dependencies among functions through hierarchical trees or network-based mechanisms. These include hierarchical specialization (depth), recombination across existing types (horizontal linkage), and the introduction of entirely novel types (breadth). 
Another promising extension involves accounting for function obsolescence. Our model focuses on the addition of individuals and functions, but individuals also leave organizations and cities, just as genes can naturally be removed from genomes through evolution. Entire functions can also become obsolete, such as the obsolescence of switchboard operators who manually connected telephone calls. Although the mechanisms of agent removal and function obsolescence are still areas of active research \cite{lee2025synthesis,saavedra2008asymmetric}, it could be useful for future research to consider the removal of individuals and functions in complex systems and how it would affect diversification and specialization. Lastly, our model treats the introduction of new functions as a single process, without distinguishing between the creation of globally novel functions and the adoption of ones that exist in other entities. In both biological and social systems, new functions arise within an entity and spread through diffusion—for example, via horizontal gene transfer in biological systems \cite{arnold2022horizontal} and imitation in social systems \cite{dimaggio1983iron}---reflecting a shared innovation–diffusion dynamic. Because the adoption of functions that exist in other organisms is likely far more common than the creation of globally new ones, it may dominate the observed empirical patterns. Future work could explicitly separate these processes to assess their relative contributions.

Our work makes several contributions to the literature. First, in contrast to the prevalent trend of studying diversification and specialization within specific system types, our work presents an integrated analysis across a diverse array of systems, encompassing both biological (bacteria) and social systems (companies, government agencies, universities, cities). This approach identifies commonalities across many biological and social systems, while also identifying notable exceptions. It provides testable hypotheses to explore patterns in other systems. Second, while previous models explained power-law or logarithmic scaling relationships using distinct formulations and assumptions, we develop a general model that explains both types of scaling within one unifying framework. Third, by estimating parameters in the framework of this mathematical model, we can quantify common characteristics across complex systems and evaluate how systems differ from each other. 

\section*{Code availability}
The data and code necessary to replicate the results of this study are available at \\\href{https://github.com/jamesholehouse/Calibrations-for-Function-Diversity}{https://github.com/jamesholehouse/Calibrations-for-Function-Diversity}.

\section*{Author contributions}
V.C.Y. and J.H. contributed equally to this work. Data gathering and empirical analysis: V.C.Y., J.I.A.; Model development, simulation, and calibration: J.H., S.R.; Writing – Initial Draft: V.C.Y, J.H.; Writing – Review \& Editing: All authors; Conceptualization: All authors.

\section*{Acknowledgments}
The authors would like to acknowledge the support of the National Science Foundation Grant Award Number EF--2133863. We would like to thank the UNM Center for Advanced Research Computing, in particular, Matthew Fricke, supported in part by the National Science Foundation, for providing the research computing resources used in this work. H.Y.~acknowledges the Emergent Political Economies grant from the Omidyar Network, the NRF Global Humanities and Social Sciences Convergence Research Program (2024S1A5C3A02042671) and the support from the Institute of Management Research at Seoul National University. J.H.~acknowledges the support of a Lou Schuyler grant from the Santa Fe Institute.

\printbibliography

\end{document}